\documentclass[aps,prb,onecolumn,preprint,superscriptaddress]{revtex4}
\usepackage{graphicx}

\begin{document}

\title{Observation of scalable and deterministic multi-atom Dicke states in an atomic vapor}


\author{Shaogang Yu}
\affiliation
{Department of Physics, Florida International University, Miami, Florida 33199, USA}

\affiliation
{State Key Laboratory of Magnetic Resonance and Atomic and Molecular Physics, Wuhan Institute of Physics and Mathematics, Chinese Academy of Sciences, Wuhan 430071, China}

\affiliation
{University of Chinese Academy of Sciences, Beijing 100049, China}

\author{Michael Titze}
\affiliation
{Department of Physics, Florida International University, Miami, Florida 33199, USA}

\author{Yifu Zhu}
\affiliation
{Department of Physics, Florida International University, Miami, Florida 33199, USA}

\author{Xiaojun Liu}

\affiliation
{State Key Laboratory of Magnetic Resonance and Atomic and Molecular Physics, Wuhan Institute of Physics and Mathematics, Chinese Academy of Sciences, Wuhan 430071, China}

\author{Hebin Li}
\email[Email address: ]{hebin.li@fiu.edu}
\affiliation
{Department of Physics, Florida International University, Miami, Florida 33199, USA}

\renewcommand\thetable{\arabic{table}}

\begin{abstract}
Dicke state, a coherent state of multiple particles, is fundamentally responsible for various intriguing collective behaviors of many-body systems. Experimental access to Dicke states with a scalable and deterministic number of particles is essential to study how many-body properties depend on the particle number. We report the observation of Dicke states consisting of two, three, four, five, six, and seven atoms in an atomic vapor. Quantum coherences between the ground state and multi-atom states are created and detected by using optical two-dimensional coherent spectroscopy. The signal originated from multi-atom states is manifested as correlation between the multi-quantum coherence and the emission signal, providing direct and unambiguous detection of Dicke states. The manipulation of deterministic atomic Dicke states has possible implications in quantum information processing and fundamental many-body physics.
\end{abstract}

\maketitle
Behaviors of many-body systems cannot always be explained by a simple extrapolation of individual properties\cite{Anderson1972}, but rather are results of a collective state of multiple particles. In 1954, Dicke introduced \cite{Dicke1954} a coherent collection of atoms in which $n$ atoms act collectively as one big atom, forming so called Dicke states. The cooperative spontaneous emission from a Dicke state is known as superradiance \cite{Dicke1954,PhysRevLett.30.309} whose intensity is proportional to $n^2$ instead of $n$. Interest in Dicke states has been revived recently due to studies of single-photon supperradiance \cite{Scully2009,PhysRevLett.96.010501}, collective lamb shift \cite{PhysRevLett.102.143601,Friedberg2008,Rohlsberger2010}, Dicke quantum phase transition \cite{Baumann2010,Nataf2010,Hamner2014}, and multi-particle entanglement\cite{Sackett2000, Haffner2005, Monz2011, Kiesel2007,Wieczorek2008,Prevedel2009,Wieczorek2009,Yao2012,Wang2016a,Barends2014,Song2017}. As a many-body model system with exact solutions \cite{PhysRev.170.379}, Dicke states can provide unique insights into how many-body properties scale with the number of particles. Experimentally, it requires the capability to prepare Dicke states of a scalable and deterministic number of particles, which has been demonstrated with trapped ions \cite{Sackett2000, Haffner2005, Monz2011}, entangled photons \cite{Kiesel2007,Wieczorek2008,Prevedel2009,Wieczorek2009,Yao2012,Wang2016a}, and superconducting qubits \cite{Barends2014,Song2017}. However, realizing Dicke states with a few neutral atoms/molecules is challenging and the collective excitation of neutral atoms/molecules has been limited to either two particles \cite{Hettich2002,Gaetan2009a,Dai2012,Gao2016} or a large ensemble\cite{PhysRevLett.30.309, Baumann2010,Hamner2014}.

Here we report the creation and detection of two-, three-, four-, five-, six-, and seven-atom Dicke states in an atomic vapor by using optical two-dimensional coherent spectroscopy (2DCS). Through a multi-photon process, femtosecond excitation pulses correlate multiple atoms and generate a multi-quantum coherence between the ground state and the multi-atom excited state. The resulting multi-quantum coherence is subsequently down-converted by excitation pulses to a single-quantum coherence which radiates a signal. The number of correlated atoms is determined by the order of multi-photon process. The nonlinear excitation up to 14th-order provides access to Dicke states of up to seven atoms. We made a significant advance in 2DCS technique to realize seven-quantum 2D spectroscopy for the first time. The obtained 2D spectra display the correlation of multi-quantum coherence and emission signal, allowing unambiguous detection on selective Dicke states. We further found that the decoherence rate increases linearly with the number of atoms, confirming the signature cooperative behavior of Dicke states.

\begin{figure}[thb]
  \centering
  \includegraphics[width=\textwidth]{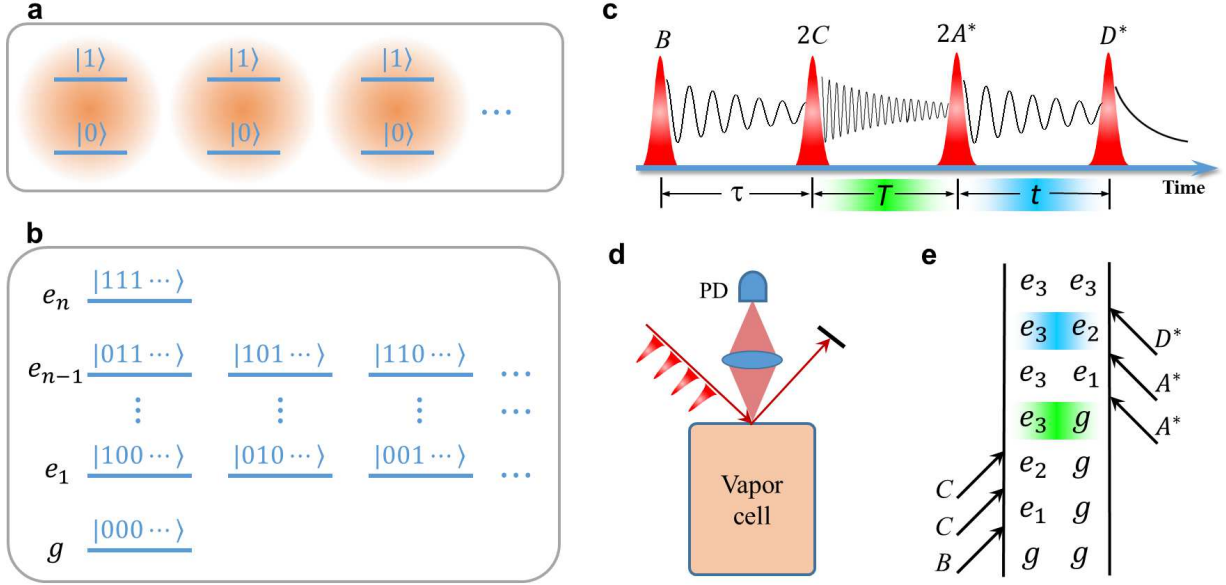}
  \caption{Experimental scheme to observe multi-atom Dicke states. \textbf{a}, Energy level diagrams of $n$ individual two-level atoms. \textbf{b}, Energy level diagram of Dicke states of $n$ two-level atoms. \textbf{c}, Excitation pulse sequence to observe three-atom Dicke states. \textbf{d}, Experimental schematic. The fluorescence signal is recorded by a photo detector (PD). \textbf{e}, Double-sided Feynman diagram showing one of the eight excitation pathways.}
  \label{fig:figure1}
\end{figure}

Consider a system of $n$ atoms with two energy levels, as shown in Fig. \ref{fig:figure1}a, each atom can be treated separately and has two states $|0\rangle$ and $|1\rangle$. In the joint basis of $n$ correlated atoms, the system as a whole can be described by using Dicke states \cite{Dicke1954} as illustrated in Fig. \ref{fig:figure1}b. The ground state $g$ has all atoms in state $|0\rangle$. The first excited state $e_1$ has one atom (but not known which one) excited and others in state $|0\rangle$. Each next higher-energy excited state has one more atom excited until all $n$ atoms are excited in state $e_n$. For $n$ correlated atoms, it is possible to generate a coherence $\rho_{ge_n}$ between states $g=|000\cdots\rangle$ and $e_n=|111\cdots\rangle$, which is known as $n$-quantum coherence. To observe the $n$-atom Dicke state $e_n$, we measure the corresponding $n$-quantum coherence $\rho_{ge_n}$ which oscillates at $n$ times the transition frequency $\omega_{10}$ of a single atom. This approach eliminates the ambiguity in detecting spontaneous emission where signals from all manifolds of the ladder have the same frequency $\omega_{10}$. It also ensures that the involved state is a coherent collection of atoms which is required for generating $n$-quantum coherence.

Multi-quantum coherence can be created and detected by using optical two-dimensional coherent spectroscopy (2DCS), which has been demonstrated to measure two- or three-quantum coherence in atoms \cite{Dai2012,Gao2016} and semiconductor quantum wells \cite{Turner2010,Karaiskaj2010,Stone2009a}. We implemented double-quantum 2DCS of atoms \cite{Dai2012,Gao2016} in a collinear setup based on acousto-optic modulators (AOMs) \cite{Nardin2013,Tekavec2007} and further extended the technique to perform multi-quantum 2DCS (see Methods for detail). The experiment has a remarkable sensitivity to detect the signal from a 14th-order nonlinear process and realize seven-quantum 2DCS.
To simplify the discussion, here we use three-quantum 2DCS as an example to briefly describe the experiment. The pulse sequence shown in Fig. \ref{fig:figure1}c is used. Four pulses  $B$, $C$, $A^*$, and $D^*$ copropagate in one beam and are incident on the vapor cell window in that order, as shown in Fig. \ref{fig:figure1}d. The time delays between pulse pairs $BC$, $CA^*$, and $A^*D^*$ are defined as $\tau$, $T$, and $t$, respectively. Each pulse is phase-modulated by an AOM at a specific frequency $\omega_i$ ($i=A, B, C, D$). For three-quantum coherence, pulses $C$ and $A^*$ each act twice and pulses $A^*$ and $D^*$ are considered conjugated. The signal due to the excitation of this particular pulse sequence (i.e. $B, C, C, A^*, A^*, D$) is modulated at the frequency $\omega_{3Q}=\omega_B-\omega_D+2(\omega_C-\omega_A)$, thus it can be isolated by lock-in detection using $\omega_{3Q}$ as the reference frequency. We are interested in the eight processes (Fig. S1 in Supplementary Information) that involve three-quantum coherence $\rho_{g{e_3}}$. One of them is illustrated by the double-sided Feynman diagram in Fig. \ref{fig:figure1}e. Quantum coherences between the ground state $g$ and partially excited states $e_{1,2}$ are generated by the first two fields during this process. A three-quantum coherence $\rho_{g{e_3}}$ is created by the third field and evolves during the time period $T$. It is subsequently converted into a single-quantum coherence which evolves during the delay $t$. In our experiment, the fluorescence signal is recorded as a function of $T$ and $t$. Fourier transforming the time-domain signal generates a 2D spectrum with two frequency dimensions $\omega_T$ and $\omega_t$ corresponding to $T$ and $t$, respectively. The 2D spectrum displays the correlation between the dynamics during $T$ and $t$ as a spectral resonance on the diagonal line $\omega_T=3\omega_t$. The signal relies on the existence of three-quantum coherence and interatomic interactions that lead to an incomplete cancellation of contributions from the possible processes \cite{Dai2012,Gao2016}. This experiment can be generalized to measure $n$-quantum coherence for $n$-atom Dicke states by applying $C$ and $A^*$ pulses $(n-1)$ times each and selecting the signal at the frequency $\omega_{nQ}=\omega_B-\omega_D+(n-1)(\omega_C-\omega_A)$. In this case, $n$-quantum coherence $\rho_{ge_n}$ is created and evolves in the time period $T$. The resulting 2D spectrum correlate the $n$-quantum coherence in $T$ with the single-quantum coherence in $t$. The spectral peak should have a $n$-quantum frequency that is $n$ times the single-quantum frequency. Therefore, $n$-quantum 2DCS provides a specific detection of $n$-atom state.

The experiment was performed on a potassium (K) atomic vapor contained in a glass reference cell (see Methods for detail). A femtosecond (fs) oscillator provides excitation pulses with a duration of $\sim$150 fs at a 76-MHz repetition rate. The laser wavelength is tuned to cover both $D_1$ (389.29 THz) and $D_2$ (391.02 THz) transitions of K atoms. At this wavelength, the two-photon energy does not match any single-atom energy states. The closest higher-lying states are $5P$ and $4D$ at frequencies 740.81 and 821.36 THz, respectively, both well outside the spectral range of the double laser frequency. Thus no two-quantum signal is expected from single-atom states.

\begin{figure}[htb]
  \centering
  \includegraphics[width=0.66\textwidth]{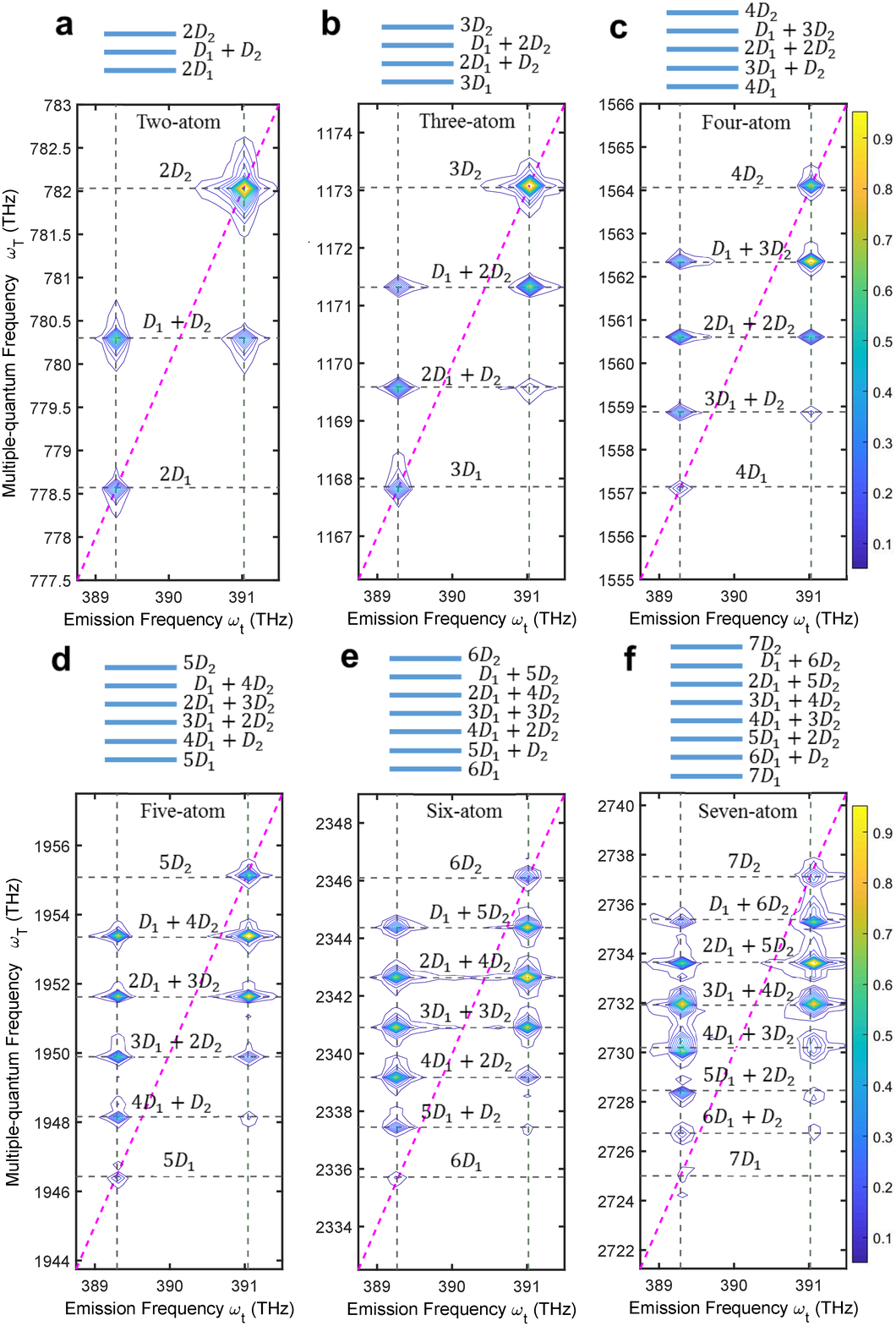}
  \caption{Multi-quantum 2D spectra of a K vapor. The spectra reveal Dicke states of \textbf{a}, two atoms, \textbf{b}, three atoms, \textbf{c}, four atoms, \textbf{d}, five atoms, \textbf{e}, six atoms, and \textbf{f}, seven atoms. In each case, the energy level diagram illustrates possible energies when all atoms in the Dicke state are excited into $D_1$ or $D_2$. The multi-quantum frequencies of each spectrum match these energies, while the emission frequencies are at $D_1$ and $D_2$ lines.}
  \label{fig:figure2}
\end{figure}

Considering both $D_1$ and $D_2$ singly excited states, the Dicke states of K atoms have multiple possible energies that are combinations of $D_1$ and $D_2$. For two-atom states, the doubly-excited state energies can be $2D_1$ (two atoms in $D_1$), $2D_2$ (two atoms in $D_2$), and $D_1+D_2$ (one atom in $D_1$ and other in $D_2$). All these two-atom states are within the spectrum of the double laser frequency and can give rise to signals in two-quantum 2D spectra. As shown in Fig. \ref{fig:figure2}a, we obtained a two-quantum 2D spectrum of a K vapor at a number density of $N=5.32 \times 10^{13}$ cm$^{-3}$. The spectrum was measured by scanning 10 ps in both $T$ and $t$ time delays and Fourier-transforming the signal in both dimensions. So the frequency resolution is 0.1 THz and the hyperfine structures are not resolved. The spectral amplitude is plotted as contour lines with the maximum value normalized to one. The vertical axis $\omega_T$ represents the two-quantum frequency that is associated with two-quantum coherences during the time period $T$. The horizontal axis $\omega_t$ is the single-quantum frequency of single-quantum coherences during $t$. The spectrum features four spectral peaks. Their two-quantum frequencies are $2D_1$, $D_1+D_2$, and $2D_2$, matching the energies of two-atom doubly excited states. The single-quantum frequency is $D_1$ ($D_2$) if the two-quantum frequency is $2D_1$ ($2D_2$) so the two corresponding peaks are located on the $\omega_T=2\omega_t$ diagonal line. If the two-quantum frequency is $D_1+D_2$, the single-quantum frequency can be either $D_1$ or $D_2$, resulting in two off-diagonal peaks. The two-quantum coherence between the ground state and two-atom doubly excited states relys on the existence of two-atom Dicke states. The signal in 2D spectrum also requires interaction between the two atoms that breaks the symmetry to avoid a complete cancellation of possible excitation pathways \cite{Dai2012}. However, the interatomic interaction is weak and no energy shift is observed within the frequency resolution in our experiment. Therefore, the two-quantum 2D spectrum of K is a direct consequence of the two-quantum coherence due to the collective state of two correlated, weakly interacting atoms, providing direct evidence of two-atom Dicke states.

\newpage

Similarly, $n$-atom Dicke states of K can be detected by using $n$-quantum 2D spectroscopy. For the case of three K atoms, the triply excited states have four energies $3D_1$, $2D_1+D_2$, $D_1+2D_2$, and $3D_2$, as shown in Fig. \ref{fig:figure2}b. The resulting three-quantum 2D spectrum includes six peaks with three-quantum frequencies that match three-atom triply excited states, as indicated by the horizontal dashed lines with corresponding labels. The peaks with a three-quantum frequency of $3D_1$ or $3D_2$ are located on the $\omega_T=3\omega_t$ diagonal line and have a single-quantum frequency of $D_1$ or $D_2$, respectivley. When the three-quantum frequency has mixed contributions from $D_1$ and $D_2$, the single-quantum frequency can be both $D_1$ and $D_2$, leading to four off-diagonal peaks. For the cases of more K atoms, possible energies of the excited state and the corresponding multi-quantum 2D spectra are shown in Fig. \ref{fig:figure2}c, d, e, and f for Dicke states of four, five, six and seven atoms, respectively. The $n$-atom excited states have $n+1$ energies due to possible combinations of $n$ atoms each being in either $D_1$ or $D_2$ states. The spectra display a similar spectral pattern as the spectra for two- and three-atom Dicke states. The multi-quantum frequencies in the vertical direction match energies of the corresponding multi-atom excited states, while the single-quantum frequencies are $D_1$ and $D_2$. The two peaks involving only $D_1$ or $D_2$ are on the corresponding diagonal line $\omega_T=n\omega_t$ in each spectrum, while other peaks are off-diagonal. The spectral pattern in the $n$-quantum spectrum is a direct result of $n$-atom Dicke states. In each 2D spectrum, the maximum signal amplitude is normalized to one. The relative strength of spectral peaks is determined by the dipole moments of $D_1$ and $D_2$ transitions and the laser spectral shape. The multi-quantum 2D spectra in Fig. \ref{fig:figure2} are the observation of Dicke states consisting of a scalable and deterministic number of atoms up to seven. The Dicke states with a specific number of atoms can be deterministically selected by using proper multi-quantum 2DCS, allowing possibilities to study the dependence of many-body properties on the number of atoms.

\begin{figure}[thb]
  \centering
  \includegraphics[width=0.4\textwidth]{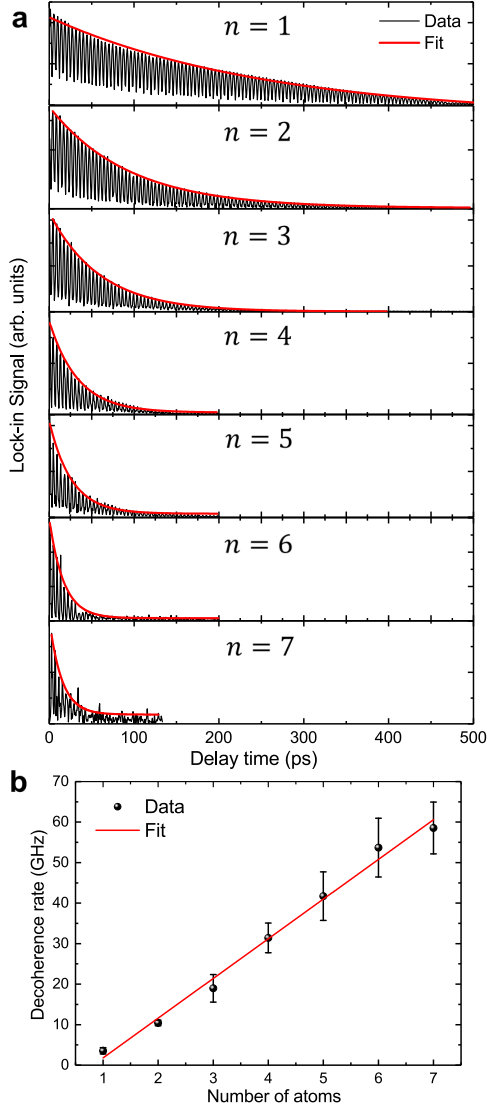}
  \caption{Decoherence dynamics of multi-atom Dicke states. \textbf{a}, The measured dynamics of $n$-quantum coherences for $n=1$ to $7$. The black lines are data and the red lines are fits to an exponential decay fucntion. \textbf{b}, The dependence of the extracted decoherence rate on the number of atoms. The black dots are data and the red line is a linear fit.}
  \label{fig:figure3}
\end{figure}

A signature collective behavior of Dicke states is superradiance \cite{Dicke1954,PhysRevLett.30.309}. The decay rate in this cooperative spontaneous emission of $n$ atoms scales as $n\gamma$ with $\gamma$ being the spontaneous decay rate of a single atom. Here we study the decoherence rate of $n$-quantum coherence, instead of the decay rate, associated with $n$-atom Dicke states and its dependence on the number of atoms. In our experiment, the $n$-quantum ($n>1$) coherence evolves during the time period $T$. To measure the decoherence dynamics of $n$-quantum coherence, we scan $T$ while keeping $\tau$ and $t$ fixed. The signal originated from $n$-quantum coherence is selected by lock-in detection at the corresponding reference frequency, similar to the detection method in multi-quantum 2DCS. The experiment is sightly different for the single-quantum coherence which is measured by scanning $\tau$ under the excitation of the single-quantum rephasing pulse sequence\cite{Nardin2013}. The measured dynamics of $n$-quantum coherence are shown in Fig. \ref{fig:figure3}a for $n=1$ to $7$, where the lock-in signal is plotted as a function of the scanned delay time. The signal should oscillate at a period of 578 fs due to the beating between $D_1$ and $D_2$, which is under sampled at a step size of 667 fs in our measurements. The peak amplitude of the signal is fit to an exponential decay function $Ae^{-\Gamma T}$, where $A$ is the amplitude and $\Gamma$ is the decoherence rate. The extracted decoherence rates are plotted as black dots in Fig. \ref{fig:figure3}b for different number of atoms. The error bars represent the standard deviations of multiple measurements. The decoherence rate increases with $n$ and can be fit to a linear function (red line), suggesting that the decoherence rate scales linearly with the number of atoms $n$. This scaling dependence can be understood by considering the decoherence processes in atomic vapor. The primary contributions to the coherence loss is the population decay and the interatomic collision. The population decay rate of $n$-atom state scales with $n$. The collision probability of $n$ atoms is $n$ times the probability of a single atom colliding with others under the same temperature and number density. So the decoherence rate due to interatomic collision is $n\Gamma^*$ with $\Gamma^*$ being the collision decoherence rate of a single atom. Considering the two decoherence processes, the overall decoherence rate is $\Gamma=n(\gamma /2+\Gamma^*)$. For K atom the $D$-lines have a lifetime of 26 ns, we can extract $\Gamma^*=9.78$ GHz from the fitting. The dependence of decoherence rate on $n$ further confirms that the observed $n$-quantum coherence is indeed a cooperative property of $n$-atom Dicke states opposed to individual atoms.

\begin{figure}[thb]
  \centering
  \includegraphics[width=0.6\textwidth]{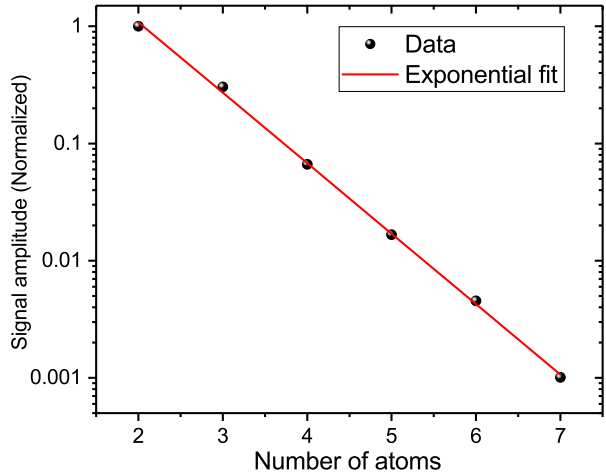}
  \caption{The signal strength of n-quantum 2D spectra. The signal amplitude is plotted as a function of the number of atoms. The black dots are the data and the red line is an exponential fit.}
  \label{fig:figure4}
\end{figure}

In our experiment, the formation of Dicke states is a result of a small number of correlated atoms due to a coherent excitation. It is possible to have simultaneously Dicke states with different number of atoms at a sufficiently high atomic density. We performed density dependence measurements of multi-quantum 2DCS (from $N=4.81\times10^8$ to $5.32\times10^{13}$ cm$^{-3}$) and found the $n$-quantum signal vanishes at a higher density for a larger $n$. The highest number of correlated atoms decreases by one for approximately one order of magnitude decrease in the atomic density. At the highest atomic density ($N=5.32\times10^{13}$ cm$^{-3}$), the $n$-quantum signals are present for $n=2$ to $7$. Comparing the overall signal strength between $n$-quantum spectra, the spectral amplitude decreases exponentially across three orders of magnitude as $n$ goes from 2 to 7, as shown in Fig. \ref{fig:figure4}. We attempted to measure an eight-quantum spectrum for eight-atom Dicke states but no signal was detected within the sensitivity of our experiment. It might be possible to observe Dicke states with eight or more atoms in further experiments with improved sensitivity and/or higher atomic density.

It has been shown that atoms/molecules in a thermalized gas can be correlated to form Dicke states and produce superradiance \cite{PhysRevLett.30.309} upon the excitation of a laser field. However, the specific Dicke state was not determined or accessible in the experiment. The significance of the multi-quantum 2DCS technique is the ability to selectively address Dicke states with a specific number of atoms even in a thermalized atomic ensemble. This capability can enable new studies of $n$-atom Dicke states in thermalized atoms which, in comparison to cold atoms, can provide a broader range of interatomic separation and may exhibit unique many-body effects and dynamics due to thermal motion. Such experiments will complement studies of many-body physics in cold atoms/molecules. On the other hand, the multi-quantum 2DCS technique can also be extended to study many-body effects in cold atoms/molecules. With the recent development of frequency-comb based 2DCS \cite{Lomsadze2017,PhysRevLett.120.233401}, one can achieve a sufficient frequency resolution to perform 2DCS on cold atoms/molecules and resolve hyperfine structures. Moreover, $n$-atom Dicke states can support quantum entanglement. It might be possible to enhance the $n$-quantum coherence to realize an entangled state so that the generated $n$-atom Dicke states can be used for quantum information processing.

In summary, we implemented optical multi-quantum 2DCS on a K vapor and obtained $n$-quantum 2D spectra with $n$ up to seven. In each spectrum, the spectral peaks match exactly the corresponding Dicke states in the multi-quantum frequency. In addition, the decoherence rate of $n$-quantum coherence was measured and found to increase linearly with the number of atoms $n$ due to the cooperative behavior of Dicke states. The $n$-quantum 2D spectra and the decoherence rates confirm the creation and detection of Dicke states with a scalable and deterministic number of atoms up to seven atoms in an atomic vapor. Access to these Dicke states will enable new measurements of many-body properties to gain insights into their dependence on the number of atoms. The technique of multi-quantum 2DCS can also be used to study other systems such as cold atoms/molecules and entangled qubits. Overall, the observation of $n$-atom Dicke states along with the technique of multi-quantum 2DCS has potential implications in the fields of fundamental many-body physics and quantum information science.

\parskip 12pt

\noindent \textbf{Methods}

\begin{figure}[thb]
  \centering
  \includegraphics[width=0.9\textwidth]{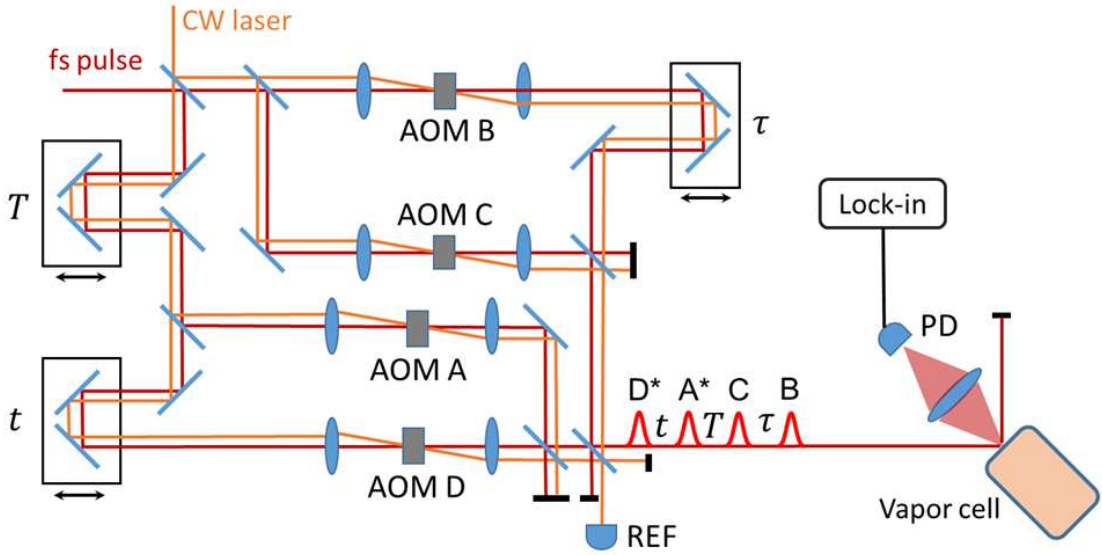}
  \caption{Schematic of the collinear two-dimensional coherent spectroscopy setup. The setup splits a femtosecond (fs) pulse into four pulses and the delays between the pulses are controlled by three delay stages. A continuous-wave (CW) laser goes through the same optics and provides an interferometric reference for lock-in detection. Each fs pulse is modulated by an acousto-optic modulator at a specific frequency. The four pulses are incident on the window of a vapor cell and the generated fluorescence signal is recorded by a photodetector (PD). The signal is demodulated by a lock-in amplifier at a proper reference frequency that is produced by mixing CW laser beating signals registered by a photodetector (REF).}
  \label{fig:figure5}
\end{figure}

\textbf{Collinear two-dimensional coherent spectroscopy setup}
A collinear optical 2-dimensional coherent spectroscopy (2DCS) setup based on acousto-optic modulators (AOMs) was implemented to perform multi-quantum 2DCS. The schematic of the setup is shown in Fig. \ref{fig:figure5}. Two nested Mach-Zehnder interferometers split a femtosecond laser pulse into four pulses labeled $B$, $C$, $A^*$, and $D^*$. Each pulse is phase-modulated by an AOM at a specific frequency $\omega_i$ ($i=A,B,C,D$). The frequencies used in our experiments are $\omega_A=80.1070$ MHz, $\omega_B=80.0173$ MHz (except for the case of seven atoms where $\omega_B=80.0193$ MHz), $\omega_C=80.1040$ MHz, and $omega_D=80.0000$ MHz. The pulses arrive at the sample in a sequence as $B$, $C$, $A^*$, and $D^*$. The time delays are defined as $\tau$ between $B$ and $C$, $T$ between $C$ and $A$, and $t$ between $A^*$ and $D^*$. The three delays are individually controlled by three delay stages. The pulse sequence is incident on the window of a vapor cell and the generated fluorescence signal is collected by a lens and recorded by a photodetector (PD). A continuous wave (CW) laser goes through the same optical path and the AOMs, but is slightly shifted vertically to be separated from the femtosecond laser beam. The CW laser beams provide beating signals at a photodetector (REF), which are processed in real time by a digital wave mixer to produce the required reference frequencies for lock-in detection. As the interferometric output of the interferometers, the reference signals derived from the CW laser also measure the beam path fluctuations which automatically correct the phase fluctuations in femtosecond laser pulses.
The output from PD is demodulated at different reference frequencies to extract the nonlinear signal for various excitation pathways defined by the pulse ordering and the phase-matching condition. For the given pulse sequence ($B$, $C$, $A^*$, and $D^*$) and the phase-matching condition $\mathbf{k}_S=\mathbf{k}_B+\mathbf{k}_C-\mathbf{k}_A-\mathbf{k}_D$, the two-quantum signal should be detected at the reference frequency $\omega_{2Q}=\omega_B-\omega_D+\omega_C-\omega_A$. Here $\mathbf{k}_i$ ($i=S,A,B,C,D$) are the wave vectors of each field. Pules $A^*$ and $D^*$ are considered conjugated so their wave vectors and modulating frequencies are negative. The reference frequency $\omega_{2Q}$ is acquired by mixing beating frequencies $\omega_{BD}=\omega_B-\omega_D$ and $\omega_{CA}=\omega_C-\omega_A$ which can be retrieved by properly filtering the output of REF. During the experiment, the demodulated signal is recorded as a function of time delays $T$ and $t$. Fourier transforming the signal with respect to the two time delays generates a two-quantum 2D spectrum. To generalize this technique to n-quantum 2DCS, we consider pulses $C$ and $A^*$ each act $n-1$ times. The $n$-quantum signal should be detected at the reference frequency $\omega_{nQ}=\omega_{BD}+(n-1)\omega_{CA}$ which is generated by the wave mixer using $\omega_{BD}$ and $\omega_{CA}$.

\textbf{Potassium vapor cell} The potassium (K) vapor cell was purchased from Thorlabs (GC25075-K). The cell is made of borosilicate glass with a diameter of 25 mm and a length of 71.8 mm. The cell is evacuated to a pressure of $10^{-8}$ Torr and filled with a small amount of K metal.



\parskip 12pt

\noindent \textbf{Acknowledgements}
\ The authors thank T.M. Autry, S.T. Cundiff, 
and G. Nardin for useful discussions. This material is based upon work supported by the National Science Foundation under Grant No. PHY-1707364.

\parskip 12pt
\noindent \textbf{Author contributions} H.L. conceived the project. S.Y and M.T. ran the experiment. S.Y. took the data and analyzed the results. H.L. and S.Y. wrote the manuscript. All authors discussed the results and commented on the manuscript at all stages.

\parskip 12pt
\noindent \textbf{Competing Interests}
\ The authors declare that they have no competing financial interests.

\parskip 12pt
\noindent \textbf{Correspondence}
\ Correspondence and requests for materials should be addressed to H.L.~(email: hebin.li@fiu.edu).

\newpage

\noindent \textbf{Supplementary information:}

\setcounter{figure}{0}
\renewcommand{\thefigure}{S\arabic{figure}}

\begin{figure}[thb]
  \centering
  \includegraphics[width=0.8\textwidth]{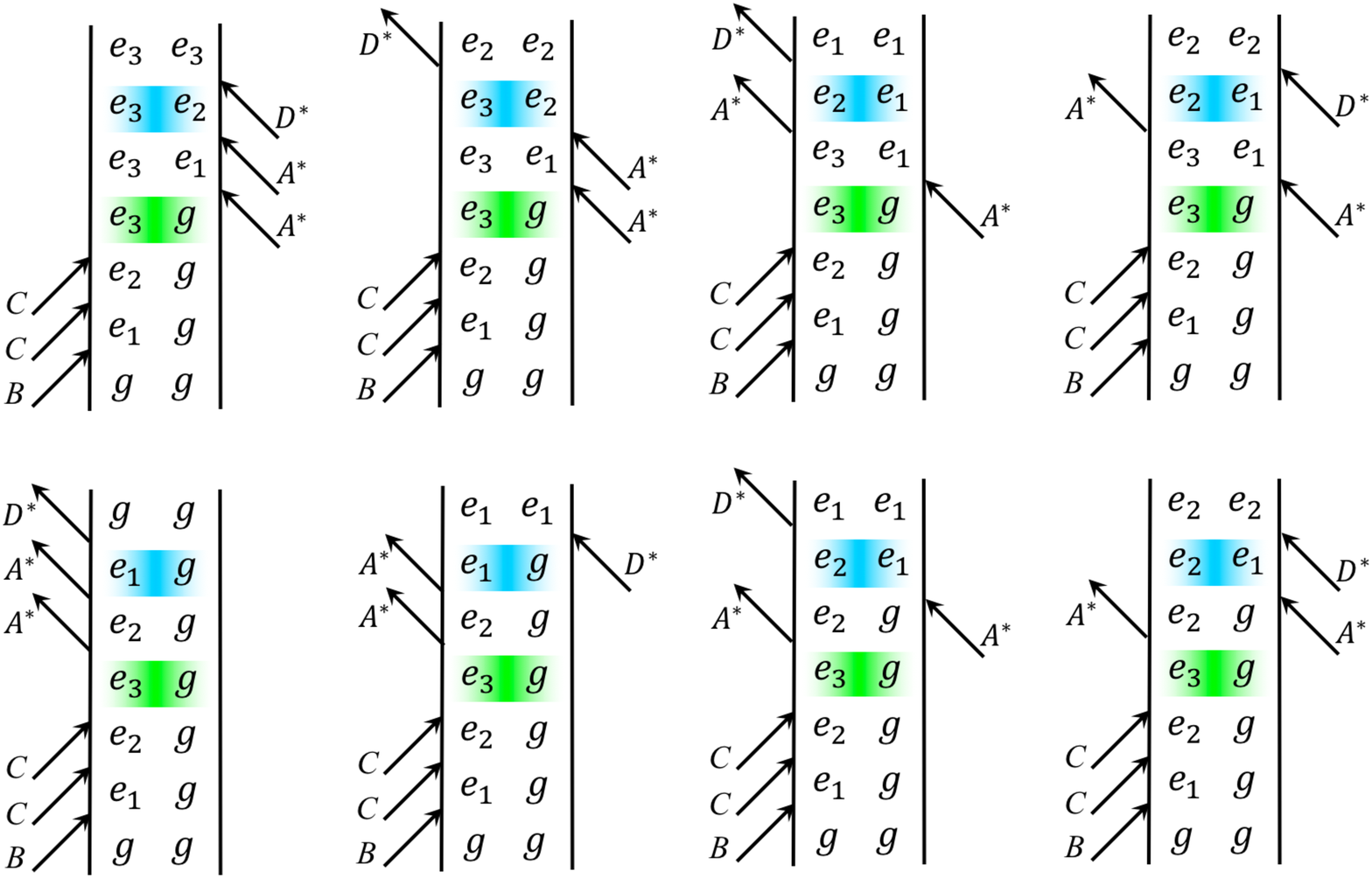}
  \caption{Double-sided Feynman diagrams of possible excitation pathways associated with three-atom states excited by the pulse sequence shown in Fig. 1c. The vertical lines indicate the time. The symbols between the vertical lines represent density matrix elements. The arrows denote acting optical fields with the excitation pointing inwards and the emission pointing outwards. The Dicke states of three two-level atoms have a ground state $g$, a first-excited state $e_1$, a second-excited state $e_2$,  and a third-excited state $e_3$. Under the excitation of the pulse sequence $B$, $C$, $C$, $A^*$, $A^*$, and $D^*$, there are eight excitation pathways that create a three-quantum coherence $\rho_{ge_3 }$ after the first three pulses. The following two pulses convert the three-quantum coherence to a single-quantum coherence and the last pulse turns the single-quantum coherence into a population. The last three pulses have eight possible combinations, resulting in eight different excitation pathways. The signal is recorded as the three-quantum coherence (green) and the single-quantum coherence (blue) evolve during the time delays T and t, respectively. The generated 2D spectrum correlates the dynamics of the three- and one-quantum coherences. }
  \label{fig:figures1}
\end{figure}

\end{document}